\documentclass[twocolumn,showpacs,preprintnumbers,amsmath,amssymb]{revtex4}
\usepackage{graphicx}
\usepackage{dcolumn}
\usepackage{bm}

\newcommand{\be}{\begin{equation}}
\newcommand{\ee}{\end{equation}}
\newcommand{\bea}{\begin{eqnarray}}
\newcommand{\eea}{\end{eqnarray}}

\begin{document}
\preprint{CPHT-RR071.0912}
\title{A Note on Hypercharge Flux, Anomalies, and U(1)s in F-theory GUTs }
\author{Eran Palti}
\affiliation{%
Centre de Physique Theorique, Ecole Polytechnique, CNRS, 91128 Palaiseau, France.}%
\begin{abstract}
We study the consequences of cancellation of cubic Abelian anomalies in F-theory GUT models that utilise hypercharge flux in the presence of additional U(1) symmetries. We show that a mixed anomaly between the hypercharge and two U(1) gauge fields is not automatically canceled in local models based on the spectral cover and therefore imposes additional constraints on local models in F-theory that have not been accounted for so far. The constraints imply that there are only two possible classes of models in F-theory GUTs which have fields on matter curves in non-complete GUT representations that are not vector-like under all U(1)s: the ones based on a $3+2$ and $2+2+1$ split. We comment on some phenomenological implications of these results for realising NMSSM-like models from F-theory GUTs.
\end{abstract}
\maketitle

\section{Introduction}

Supersymmetric grand unification of the gauge symmetries of the Standard Model remains one of the most appealing ideas in theoretical physics. One of the key motivating factors is that the representations of the MSSM form complete $SU(5)$ multiplets, with the exception of the Higgs doublets. The presence of doublets and no completing triplets, so called doublet-triplet splitting, is one of the puzzles for which string theory offers completely different, and arguably more attractive, solutions to than four-dimensional GUTs. The earliest proposal for doublet-triplet splitting was through the use of Wilson lines, but more recently it was realised that a different possibility is to turn on hypercharge flux \cite{Beasley:2008kw,Donagi:2008kj}. In $SU(5)$ F-theory GUTs this flux should be turned on along a curve which is non-trivial in the homology of the GUT surface, but which is trivial in the homology of the full Calabi-Yau. This is so that the flux does not couple to the closed string sector which would induce a Stueckelberg mass for the hypercharge. Such a flux breaks the unified gauge group from $SU(5)$ to the commutant of the hypercharge generator inside $SU(5)$ which is the SM gauge group. Doublet-triplet splitting is then achieved by an appropriate restriction of the flux to the matter curves carrying the Higgs representations. 

The fact that the hypercharge flux must be globally trivial constrains the possible spectrum that arises from it. For example the spectrum of fields that do not form complete GUT multiplets must be vector-like under the SM gauge groups since the net chirality is given by a topological index and the hypercharge flux is topologically trivial. More sophisticated geometric constraints were noted in \cite{Marsano:2009gv,Marsano:2009wr,Dudas:2009hu,Dudas:2010zb,Marsano:2010sq,Dolan:2011iu} for models that are based on the Tate form of the elliptic fibration of F-theory and its local version around the GUT brane which is described using the spectral cover \cite{Donagi:2009ra}. These are particularly stringent in the presence of additional $U(1)$ symmetries that remain after breaking
\be
E_8 \rightarrow SU(5)_{GUT} \times SU(5)_{\perp} \rightarrow SU(5)_{GUT} \times \Pi_i U(1)_i \;. \label{e8u1}
\ee 
It was pointed out in \cite{Marsano:2010sq} that the constraints on the spectrum that were deduced geometrically can be understood in terms of four-dimensional anomaly cancellation. The important point is that because the hypercharge flux can not couple to the closed string sector it can not modify the Green-Schwarz anomaly cancellation mechanism and so any anomalies in the spectrum after GUT breaking must be cancelled by the same diagrams as before GUT breaking.
Hence hypercharge flux can not induce any anomalies in the spectrum, and while this is automatic for anomalies involving the SM gauge groups because there is no net chirality induced, the anomalies involving an additional $U(1)$ are only canceled once the appropriate constraints on the spectrum are imposed. This does not mean that the $U(1)$ must not be anomalous, just that its anomalies must be proportional to any anomalies present already at the GUT level before breaking. The fact that this subtle anomaly cancellation condition is encoded automatically in the local geometry was rather satisfying and inspired confidence in the spectral cover approach.

The relevant anomalies involving the additional $U(1)$ that lead to the interesting constraints on the spectrum were the mixed anomalies present after hypercharge flux breaking
\be
{\cal A}_{SU(3)^2 - U(1)} \;,\; {\cal A}_{SU(2)^2 - U(1)} \;,\; {\cal A}_{U(1)^2_Y - U(1)} \;. \label{mixanom}
\ee
There is another anomaly present in the case of a $U(1)$ which is 
\be
{\cal A}_{U(1)_Y - U(1)^2} \;. \label{y2anom}
\ee
Within a possible Green-Schwarz counterterm to this anomaly the hypercharge flux appears directly through a Stuckelberg mass coupling and therefore, in order to remain massless, such a counterterm must vanish. Another way to see that it must vanish is that according to the discussion above it must be proportional to the GUT anomaly, but the latter vanishes automatically because of the non-Abelian GUT group ${\cal A}_{SU(5) - U(1)^2} = 0$.
Therefore for globally trivial hypercharge flux we would expect that the local geometry should impose this constraint on the spectrum. However, unlike the anomalies (\ref{mixanom}) this does not come out from all currently known constraints on the geometry. Therefore we must deduce that there are additional constraints on the local geometry of F-theory GUT models that have a (possibly anomalous and massive) $U(1)$ symmetry. The purpose of this note is to study what these constraints are and what are their consequences. We will see that these turn out to be quite restrictive.

\section{Constraints from Hypercharge Anomalies}
\label{constan}

The massless representations on a given matter curve are determined by two types of fluxes: there are GUT universal fluxes, for example along the $U(1)$ generators, and there is the hypercharge flux. Since the anomaly (\ref{y2anom}) vanishes when tracing over complete GUT representations the GUT universal fluxes can not affect it. Therefore we are only interested in the spectrum induced by the hypercharge flux which for 5 and 10 curves denoted ${\cal C}_5^j$ and ${\cal C}_{10}^i$ is given by \cite{Beasley:2008kw,Donagi:2008kj,Marsano:2009gv}
\bea
{\cal C}_5^j\;&:&\;\chi\left[\left(1,2\right)_{+\frac12}^{Q_5^j}\right]= N_5^j \;, \nonumber \\
{\cal C}_{10}^i\;&:&\;\chi\left[\left(\bar{3},1\right)_{-\frac23}^{Q_{10}^i}\right]= -N_{10}^i \;, \nonumber \\
{\cal C}_{10}^i\;&:&\;\chi\left[\left(1,1\right)_{+1}^{Q_{10}^i}\right]= N_{10}^i \;. \label{bastat}
\eea
Here $\chi$ denotes the net chirality, the hypercharge flux restriction to the curve is defined as $N_{5}^j \equiv c_1\left({\cal L}_Y^{5/6}\right) \cdot {\cal C}_{5}^j$, and $Q_{5}^j$ is the charge of the representation under the $U(1)$. Therefore the anomaly implies
\be
3 \sum_{{\cal C}_{10}^i} \left(Q_{10}^i\right)^2 N_{10}^i + \sum_{{\cal C}_{5}^j} \left(Q_{5}^j\right)^2 N_{5}^j = 0 \;. \label{anomN}
\ee
For convenience it is worth quoting in proximity the similar expressions for the anomalies (\ref{mixanom}) 
\be
\sum_{{\cal C}_{10}^i} Q_{10}^i N_{10}^i + \sum_{{\cal C}_{5}^j} Q_{5}^j N_{5}^j= 0 \;, \label{anomNmix}  
\ee
and for the constraint of no net chirality
\be
\sum_{{\cal C}_{10}^i} N_{10}^i = \sum_{{\cal C}_{5}^j} N_{5}^j = 0 \;. \label{nochiN}  
\ee
We proceed to analyse these anomalies for models where we consider the breaking (\ref{e8u1}). There are multiple possible breaking patterns depending on how the $U(1)$ factors are embedded into $SU(5)_{\perp}$. These possibilities are labeled in F-theory using the language of monodromies, with the different patterns shown in table \ref{u1emb}.
\begin{table}
\centering
\begin{tabular}{|c|c|c|}
\hline
Breaking Pattern & Number of $U(1)$s & $5$-matter curves \\
\hline
$\left[4\oplus 1\right]$ & 1 & 2 \\
$\left[3\oplus 2\right]$ & 1 & 3 \\
$\left[3\oplus 1\oplus 1\right]$ & 2 & 4 \\
$\left[2\oplus 2\oplus 1\right]$ & 2 & 5 \\
$\left[2\oplus 1\oplus 1\oplus 1\right]$ & 3 & 7 \\
$\left[1\oplus 1\oplus 1\oplus 1\oplus 1\right]$ & 4 & 10 \\
\hline
\end{tabular}
\caption{Table showing possible $U(1)$ embeddings, the number of $U(1)$s and the number of 5-matter curves.}
\label{u1emb}
\end{table}

Consider the parameterisation of the $U(1)$ embedding into $SU(5)$ in terms of 5 $U(1)$s labeled by $t^a$ with $a=1,...,5$. A general $U(1)$ is given by a combination $U(1):u_a t^a$ up to a single tracelessness constraint on the $u_a$ which ensures that the embedding is into $SU(5)$ rather than $U(5)$. Then in the case of no monodromies the charges of the matter curves under these $U(1)$s are given by
\bea
Q_{10}^i &:& t_a  \;, \label{cur1} \\
Q_{5}^j &:& -t_a - t_b , \;\; a\neq b\;, \label{cur2} 
\eea
where the charge under a $U(1)$ is given by contracting the $t^a$ and $t_a$ under the rule $t^a t_b = \delta^a_b$. Note that the number of ${\cal C}_{10}^i$ is always given by the number of $t_a$. The tracelessness constraint in this case is simply that a $U(1)$ acting on $\sum_a t_a$ gives zero charge. When monodromies are present (the non-unity factors in table \ref{u1emb}) the same formalism holds but we simply identify the appropriate $t_a$ with each other.

The additional anomaly constraint is simplest for the cases of a single $U(1)$. In the $\left[4\oplus 1\right]$ case we have 4 free flux parameters (two $5$-matter curves and two $10$-matter curves) and 4 constraints from (\ref{anomN}-\ref{nochiN}). The unique solution is that all the hypercharge flux must vanish. Similar counting for the $\left[3\oplus 2\right]$ split, with the unique traceless $U(1)=-2t^1+3t^2$, implies a one-parameter flux configuration and this is given in table \ref{u1emb2}.
\begin{table}
\centering
\begin{tabular}{|c|c|c|}
\hline
Curve & Charge & Flux \\
\hline
${\cal C}_{10}^1$ & -2 & N \\
\hline
${\cal C}_{10}^2$ & 3 & -N \\
\hline
${\cal C}_{5}^1$ & 4 & N \\
\hline
${\cal C}_{5}^2$ & -1 & -N \\
\hline
${\cal C}_{5}^3$ & -6 & 0 \\
\hline
\end{tabular}
\caption{Table showing one-parameter family of hypercharge flux configurations compatible with anomalies for the case of a $\left[3\oplus 2\right]$ split.}
\label{u1emb2}
\end{table}
In particular note that this is incompatible with the naive doublet-triplet splitting configuration for the Higgs curve assignments determined by the Yukawa couplings \cite{Marsano:2009wr} in that the flux on the up-Higgs curve ${\cal C}_{5}^3$ must vanish. Doublet-triplet splitting can be achieve in principle by coupling the Higgs triplet to a triplet on the matter curve and giving them a mass by an appropriate singlet vev that breaks the $U(1)$, however this would also induce large R-parity violating terms.

For the cases where there are multiple $U(1)$s present it is informative to do some parameter and constraint counting comparisons. Consider the case with no monodromies, then the constraint (\ref{anomNmix}) must hold for the 4 traceless linearly independent $U(1)$ combinations. However we can also forget about the tracelessness constraint and take a basis of 5 $U(1)$s labeled by $t^a$ and impose that (\ref{anomNmix}) holds for all 5 of these $U(1)$s. Naively this would seem like over constraining, but if we consider the combination $U(1):\sum_a t^a$ then (\ref{anomNmix}) is automatically implied from (\ref{nochiN}). Hence we can consider a basis of $U(1)$s given by the $t^a$. It is straightforward to check that the reasoning follows also for the cases with non-trivial monodromies. The result is that (\ref{anomNmix}) must hold for each of the $t_a$. This therefore implies a unique solution 
\be
N_{10}^a = -\sum_{{\cal C}_{5}^j} \left(Q_{5}^j\right)^a N_{5}^j \;,
\ee
where the new superscript $a$ on the charge denotes that it is with respect to the $U(1)$ given by $t^a$. On top of this we must impose one additional constraint from (\ref{nochiN}) and therefore before imposing (\ref{anomN}) the number of free parameters is given by one less than the number of $5$-matter curves. In the case of multiple $U(1)$s we must generalise (\ref{anomN}) to include fully mixed anomalies
\be
3 \sum_{{\cal C}_{10}^i} \left(Q_{10}^i\right)^A\left(Q_{10}^i\right)^B N_{10}^i + \sum_{{\cal C}_{5}^j} \left(Q_{5}^j\right)^A\left(Q_{5}^j\right)^B N_{5}^j = 0 \;, \label{anomNf}
\ee
where the indices $A$ and $B$ range over any linearly independent basis of the $U(1)$s. Therefore the number of constraints is $1,3,6,10$ for $1,2,3,4$ $U(1)$s respectively. Comparing this with the number of $5$-matter curves given in table \ref{u1emb} we see that only the case $\left[2\oplus 2\oplus 1\right]$ allows for, a one-parameter family of, non-vanishing solutions which is presented in table \ref{u1emb3} \footnote{Actually there is additional structure in this case to do with the choice of monodromy action being ${\mathbb Z}_2$ or ${\mathbb Z}_2\times{\mathbb Z}_2$, and in the latter case there are six $5$-matter curves \cite{Heckman:2009mn,Dudas:2009hu} implying a two-parameter family of solutions. This is just a simple generalisation of table \ref{u1emb3} with ${\cal C}_{5}^5$ splitting into two curves and accordingly $N_5^5$ splitting into two fluxes that sum to zero.}. 
\begin{table}
\centering
\begin{tabular}{|c|c|c|c|c|c|}
\hline
Curve & Charge & Flux & Curve & Charge & Flux \\
\hline
${\cal C}_{10}^1$ & $\left(1,0\right)$ & -N & ${\cal C}_{5}^2$ & $\left(0,-2\right)$ & 0 \\
\hline
${\cal C}_{10}^2$ & $\left(0,1\right)$ & N & ${\cal C}_{5}^3$ & $\left(1,2\right)$ & -N\\
\hline
${\cal C}_{10}^3$ & $\left(-2,-2\right)$ & 0 & ${\cal C}_{5}^4$ & $\left(2,1\right)$ & N\\
\hline
${\cal C}_{5}^1$ & $\left(-2,0\right)$ & 0 & ${\cal C}_{5}^5$ & $\left(-1,-1\right)$ & 0\\
\hline
\end{tabular}
\caption{Table showing one-parameter family of hypercharge flux configurations compatible with anomalies for the case of a $\left[2\oplus 2\oplus 1\right]$ split. The particular charges given are with respect to $U(1)_1=t_1-2t_5$ and $U(1)_2=t_3-2t_5$ where the monodromy is taken to act by identifying $t_1 \leftrightarrow t_2$ and $t_3 \leftrightarrow t_4$. Note that the hypercharge flux must restrict trivially to any candidate up-Higgs curves.}
\label{u1emb3}
\end{table}

\section{GUT Anomalies and the Local Geometry}
\label{constangut}

In the previous section we derived (\ref{anomN}) from anomalies induced by trivial hypercharge flux. In this section we will argue that the same constraint can be seen by considering a different set of anomalies already before GUT breaking. Since the spectral cover models the $U(1)$ branes as part of the gauge group over $S_{GUT}$ we can also consider turning on globally trivial $U(1)$ flux and study the chirality and anomalies this induces in the spectrum. The general local model should be consistent with such a possibility. The trivial $U(1)$ flux can induce chirality with respect to the $SU(5)$ gauge group but summing over the {\bf 5} and {\bf 10} representations there should be not net chirality, which is indeed the case as enforced by non-Abelian anomaly cancellation. The chirality induced by $U(1)$ flux is proportional to the charge of the matter curve under this $U(1)$, just as in the hypercharge case. Therefore the mixed ${\cal A}_{SU(5)^2-U(1)}$ anomaly, which must vanish, is quadratic in the $U(1)$ charges, and it is simple to check that it exactly reproduces (\ref{anomN}) where now the $N_i$ are restrictions of the trivial $U(1)$ fluxes. This is a nice cross-check on the constraints and supports the view that the constraints are not dependent on the fluxes but rather are statements regarding the possible globally trivial components of the matter curves.

Unlike the constraints (\ref{anomNmix}) and (\ref{nochiN}), the constraint (\ref{anomN}) does not come out from the type of local geometry or spectral cover analysis presented in \cite{Marsano:2009gv,Marsano:2009wr,Dudas:2009hu,Dudas:2010zb,Marsano:2010sq,Dolan:2011iu,Palti:2012aa} for example. These analyses calculated the homology classes of the matter curves using the spectral cover or the local limit of a Tate model. Over all the matter curves the distinct components that a globally trivial flux could possibly restrict to were limited: essentially there was one homology class which could be globally trivial for every $U(1)$ factor. The possible trivial components determined above are not compatible with the homology calculations of the $\left[3\oplus 2\right]$ and $\left[2\oplus 2\oplus 1\right]$ cases presented in \cite{Marsano:2009wr} and \cite{Dudas:2010zb}. However it was pointed out in \cite{Dolan:2011iu} that in fact the spectral cover allows for more general restrictions of trivial flux than given in the previous studies due to an extra freedom in solving the tracelessness constraint. Indeed it was claimed that all configurations compatible with (\ref{anomNmix}) and (\ref{nochiN}) could be constructed, at least at the local level, which means that the models presented in section \ref{constan} could potentially be realised in F-theory. 

It would be very interesting to understand how the constraints on the geometry from anomalies can arise in F-theory global models. This would require a good understanding of $U(1)$s in global models (see \cite{Grimm:2010ez} for some recent work) and of how anomaly cancellation manifests in the geometry \cite{moregrimmtocome}.

\section{Some Phenomenological Implications}

The analysis presented has revealed that the full anomaly constraints are very restrictive on GUT models which utilise hypercharge flux to affect the matter spectrum with respect to any additional $U(1)$ symmetries, whether themselves are anomalous or not. The small set of possible models can be studied quite exhaustively \cite{cdp2}. 

It is important to emphasise that the lack of net restriction of hypercharge flux to the matter curves does not imply that it can not induce doublet-triplet splitting, just that it can not induce any non-GUT structure with respect to any $U(1)$ symmetry, ie. the Higgs fields must be vector-like under all $U(1)$s.

We also note that breaking to $SU(5)$ with flux rather than Higgsing can add additional charged bulk matter which could be useful in weakening these constraints.

For now we just make some comments regarding implications for NMSSM-like constructions. One of the attractive motivations for the NMSSM is that the Higgs superpotential mass term is forbidden by some $U(1)$ symmetry which does allow for a $X H_u H_d$ operator instead, with $X$ being a gauge singlet. If we wish to realise this, and also doublet-triplet splitting with hypercharge flux, then an immediate implication, already from (\ref{anomNmix}) is that we must have additional exotic states in the spectrum whose mass is protected by the same $U(1)$ that is broken by $X$. Therefore they must also lie around the TeV scale. Such models have two fairly universal properties in the constrained framework of section \ref{constan}. The first is that since there is no flux on the possible up-Higgs curves, defined as curves that due to the monodromy can allow for an up-type Yukawa coupling involving the same $10$-matter curve, there are exotic triplets in the spectrum with the same $U(1)$ charges as the up-Higgs. Quite generally such triplets induce dimension 5 proton decay operators with coefficients inversly proportional to the vev of the singlet that lifts the exotics, and it should be checked on a model by model basis that this can be made consistent with exprimental limits.

The second feature is universal to all such models with exotics present at the TeV scale: since the hypercharge flux must restrict non-trivially to the $10$-matter curves sector the only possibility of forming complete GUT multiplets from the exotics is to have at least 2 vector pairs of $10$s and 1 vector pair of $5$s. This is incompatible with perturbative gauge coupling at the GUT scale. Therefore the only possibility compatible with perturbative gauge coupling unification is that they come in the combination $\left[\left(3,2\right)_{\frac16} \oplus  \left(\bar{3},1\right)_{\frac13} \oplus 2\times\left(1,1\right)_{1} \right]$. Indeed in \cite{Dudas:2010zb} this combination arose frequently, and it can also arise in the model of table \ref{u1emb2} as shown in \cite{Palti:2016kew}. This is of interest particularly because LHC results on the Higgs mass and couplings may be nicely accommodated within an NMSSM framework \cite{Ellwanger:2011aa}.

{\bf Acknowledgments}
I would like to thank Pablo Camara and Emilian Dudas for many enlightening discussions. The research of EP is supported by a Marie Curie Intra European Fellowship within the 7th European Community Framework Programme.


\end{document}